\documentclass[onecolumn,usenatbib,amsmath,amssymb,amsbsy]{mn2e}

\usepackage{natbib}
\newcommand{\be}{\begin{equation}}
\newcommand{\ee}{\end{equation}}
\newcommand{\bear}{\begin{eqnarray}}
\newcommand{\eear}{\end{eqnarray}}
\newcommand{\hn}{\hat{n}}
\newcommand{\hxi}{\hat{\xi}}

\begin{document}

\title[Lagrangian perturbation theory for rotating magnetic stars]
{Lagrangian perturbation theory for rotating magnetic stars}

\author[Glampedakis \& Andersson]{Kostas Glampedakis and Nils Andersson\\
School of Mathematics, 
University of Southampton, Southampton SO17 1BJ, United Kingdom}
 
\maketitle

\begin{abstract}
Motivated by the possibility of radiation driven instabilities in rotating magnetic stars, we
study the stability properties of general linear perturbations of a stationary and axisymmetric,
infinitely conducting perfect fluid configuration threaded by a magnetic field and surrounded by
vacuum. We develop a Lagrangian perturbation framework which enables us to formulate a strict stability criterion 
based on the notion of a canonical energy (a functional of the fluid displacement $\xi $ and its first time derivative). 
For any given choice of $\{\xi,\partial_t \xi  \} $, the sign of the canonical energy determines whether the 
configuration is stable or not at the linear level. Our analysis provides the first complete description of
the stability problem for a magnetic star, allowing for both rotation and the presence of a magnetic field in the exterior
vacuum region. A key feature of the Lagrangian formulation is the existence of 
so-called `trivial' fluid displacements, 
which do not represent true physical perturbations. In order for the stability criterion to make rigorous sense 
one has to isolate these trivials and consider only the physical `canonical' displacements. 
We discuss this problem and formulate a condition which must be satisfied by all canonical displacements.    
Having obtained a well-defined stability criterion we provide examples which 
indicate that the magnetic field has a stabilising effect on radiation driven instabilities.

\end{abstract} 
  
\maketitle

%%%%%%%%%%%%%%%%%%%%%%%%%%%%%%%%%%%%%%%%%%%%%%%%%%%%%%%%%%%%%%%%%%%%%%%%%%%%%%%%%%%%%%%%%%%%%%%%%%%%%%%%%%%%%%%

\section{Introduction}
\label{sec:intro}

The stability of a hydromagnetic configuration under the influence of linear perturbations is
a classic interdisciplinary problem whose relevance ranges from astrophysical systems to terrestrial laboratories. Accordingly, 
the  literature on the subject is vast (see \citet{mestelbook} for an up-to-date discussion of
magnetic astrophysical systems and a rich list of relevant references). 
The present paper is motivated by the interesting possibility that the emission of radiation may drive various 
stellar oscillation modes unstable. This class of instabilities has so far only been studied in detail 
for systems radiating gravitational waves
at a significant level (see \citet{NA_CQG} for a  review). A very important question concerns to what extent
electromagnetic radiation is relevant in this context. To begin addressing this question, we 
 focus on the stability of astrophysical hydromagnetic 
systems, self-gravitating fluid bodies endowed with both rotation and magnetic fields. Neutron stars  are 
prime examples of such systems, as they are the most rapidly spinning and highly 
magnetised systems known in the cosmos. 

The foundations for the modern stability theory for Newtonian fluid stars were laid by Chandrasekhar 
more than 50 years ago (see \citet{chandra} for a summary).
A major contribution to the subject was made by \citet{bernstein}  who devised a stability criterion, based on the sign of the 
potential energy, for a static hydromagnetic configuration where the unperturbed magnetic field is 
tangential at the fluid-vacuum interface. Subsequent work by \citet{rotenberg} generalised the 
analysis to stationary configurations (albeit only for fields that are confined to the fluid). 
Another key generalisation was provided by \citet{kovetz} who
considered  the static problem but allowed the magnetic field to extend into the  vacuum domain without any 
restrictions. 

For fluid configurations, an important contribution was made by \citet{ostriker}. Their stability
criterion for a rotating fluid configuration was expressed in terms of an operator which depends on the Lagrangian 
displacement $\xi$. Since this operator represents the system's potential energy, this 
criterion is identical to that of \citet{bernstein}.    
A few years later \citet{CFS}  added a crucial element that had been overlooked in the 
previous analyses. They reformulated the stability criterion in terms of the  Lagrangian 
canonical energy functional $E_c(\xi,\partial_t \xi) $, emphasising
the existence of `trivial' fluid displacements. Even though these trivial displacements
 produce no physical change in the system,
they do change the canonical energy. Consequently,  as argued by \citet{CFS}, the 
presence of trivials disables the 
stability criterion of Lynden-Bell \& Ostriker. To avoid this impasse \citet{CFS} identified the family of trivial displacements, 
and expressed the stability criterion in terms of physical `canonical' displacements. Since the late 1970s 
this stability criterion for rotating fluid configurations has been the final word on the topic. A particularly important outcome of
the \citet{CFS} work was the discovery of the so-called CFS instability by which {\it all } rotating perfect fluid
stars are secularly unstable due to the emission of gravitational radiation \citep{CFS2,NA_CQG}.

Quite surprisingly, to the best of our knowledge, there has so far been no attempt to modify the stability criteria for 
magnetic stars along the lines of \citet{CFS}.
The original \citet{bernstein} criterion is now  textbook material (see \cite{mestelbook}). Yet 
it is clear that the existence of trivial displacements may 
{\it invalidate} the analysis for rotating stars. In fact, since they are
characterised by a non-vanishing canonical energy, trivial perturbations (which are totally unphysical) might
well be qualified as being unstable. This is troublesome, and highlights the need for the problem to be revisited.  

In this paper we aim to provide a rigorous stability criterion for
rotating hydromagnetic configurations taking due account of the presence of trivial displacements.
We consider a self-gravitating perfect fluid threaded by a magnetic field, assuming a stationary equilibrium and an arbitrary 
structure for the magnetic field (extendable to the vacuum exterior). This set up combines and 
generalises previous studies \citep{bernstein,rotenberg,kovetz}. We determine the conditions that govern the family of trivial displacements 
and provide explicit conditions that need to be satisfied by physically acceptable canonical perturbations. The required
canonical energy functional is also explicitly formulated. Given these results, the hydromagnetic stability criterion
can be properly formulated and evaluated for any magnetic field configuration.  
To illustrate this, 
we apply the stability criterion to a few simple cases. In particular, we demonstrate that there are always canonical data which make
a magnetic rotating star secularly unstable, under the emission of electromagnetic and/or gravitational radiation, in the limit
of $m \gg 1 $ (assuming that the perturbations depend on the azimuthal angle as $e^{im\varphi}$).

% Another illustrating 
%test-case is the stability of a purely toroidal mode.        

%%%%%%%%%%%%%%%%%%%%%%%%%%%%%%%%%%%%%%%%%%%%%%%%%%%%%%%%%%%%%%%%%%%%%%%%%

\section{Lagrangian perturbation theory for magnetic stars}
\label{sec:theory} 

\subsection{Formulation}
\label{sec::formulation}

We consider a {\it stationary}, but otherwise arbitrary, self-gravitating perfect fluid configuration (of compact support,
surrounded by vacuum) which is threaded by a magnetic field $ B^a $. The fluid flow is described by the velocity field
$u^a$ and furthermore, ideal MHD conditions (i.e. infinite conductivity) and a one-parameter equation of state $p=p(\rho)$ 
are assumed. We will carry out the analysis in a coordinate basis, distinguishing between co- and contra-variant 
objects. This may be a somewhat unfamiliar approach to a ``standard'' Newtonian fluid dynamics problem, but
it has significant advantages (see \citet{livrev} for a recent discussion). 
Most importantly, it allows us to define Lagrangian perturbations and
discuss conservation laws in terms of the Lie derivative. The advantage of this will be clear at various points of our discussion, 
eg. when we consider the conditions at the surface of the star. As usual in this context spatial indices are raised and lowered
with the flat metric $g_{ab}$, i.e. we have $u_a = g_{ab} u^b$ etcetera.   

The coupled fluid-magnetic field system is governed by the familiar Euler and 
induction equations (denoting the convective derivative by  $D/Dt = \partial_t + u^b \nabla_b $), 
\bear
&& \frac{D u_a}{Dt} + \frac{1}{\rho} \nabla_a 
\left ( p + \frac{B^2}{8\pi} \right ) + \nabla_a \Phi -\frac{1}{4\pi \rho}
B^b \nabla_b B_a = 0
\label{euler}
\\
\nonumber \\
&& \frac{D B_a}{Dt} = (B^b \nabla_b) u_a -B_a (\nabla_b u^b) 
%&& \partial_t B_a = (B^b \nabla_b) u_a -B_a (\nabla_b u^b) 
%-(u^b \nabla_b) B_a
\label{induction}
\eear
together with the continuity equation,
\be
\partial_t \rho + \nabla_a ( \rho u^a) = 0
\label{cont}
\ee
and Poisson's equation for the gravitational potential,
\be
\nabla^2 \Phi = 4\pi G \rho
\ee 
Additional equations are Ohm's law (which determines the electric field within
the fluid),
\be
E^a = -\frac{1}{c}\,\epsilon^{abc}u_b B_c 
\label{ohm}
\ee
and, of course, the fact that $ \nabla_a B^a = 0 $. 

The stationary 'background' configuration is obtained from the above system by setting all
time derivatives to zero.  
We wish to consider an arbitrary perturbation of this background configuration.
For the perturbative quantities we choose to work in the Lagrangian sense, by means of 
a displacement vector $\xi^a $. By definition we know that,
\be
\Delta u^a = \partial_t {\xi}^a
\ee 
For the covariant component we have,
\be
\Delta u_a = \partial_t {\xi}_a + u^b\nabla_b \xi_a + u^b\nabla_a \xi_b
\ee
where we have used
\be
\Delta g_{ab} = \nabla_a \xi_b + \nabla_b \xi_a 
\ee
For an arbitrary tensor  $Q$ (suppressing indices) we have the general expression, \citep{CFS},
\be
\Delta Q = \delta Q + {\cal L}_{\xi} Q
\ee
relating Lagrangian and Eulerian perturbations. For a scalar quantity $\rho$ the Lie derivative is given by
\be
{\cal L}_\xi \rho = \xi^a \nabla_a \rho 
\ee
Meanwhile, we have
\be
{\cal L}_\xi u_a = \xi^b \nabla_b u_a + u_b \nabla_a \xi^b 
\ee
and
\be
{\cal L}_\xi u^a = \xi^b \nabla_b u^a - u^b \nabla_b \xi^a 
\ee
for co- and contravariant vectors.  
 
Our first task is to express $\Delta \rho,~\Delta p,~\Delta B^a $ in terms of the displacement $\xi^a $. 
From conservation of mass and the equation of state we readily find \citep{CFS},
\be
\Delta \rho = -\rho \nabla_a \xi^a, \qquad \Delta p = \frac{\gamma p}{\rho} \Delta \rho  
\ee
where $\gamma = (\rho/p)(\partial p/\partial \rho ) $ is the adiabatic index.

In order to find $\Delta B^a $ we can perturb the induction equation (\ref{induction}).
The procedure becomes  more elegant  if we first rewrite  this equation (with the help of
eqn.~(\ref{cont})) using the Lie-derivative,
\be
\left [\partial_t + {\cal L}_u \right ] \left ( \frac{B^a}{\rho} \right ) = 0
\label{ind_L}
\ee  
This relation shows that $B^a/\rho$ is preserved along the flow $u^a$. 
This encodes the expectation that the field lines are `frozen into' the fluid.
An attractive feature of (\ref{ind_L}) is that the $\Delta$ and  $[\partial_t + {\cal L}_u ] $ operators commute. 
Hence, we immediately have
\be
[\,\partial_t + {\cal L}_u\, ]\, \Delta \left (\frac{B^a}{\rho} \right ) = 0
\label{DB1}
\ee
In other words, the Lagrangian perturbation of $B^a/\rho$ is also {\it Lie-transported} by the flow. 
It is natural to assume that the initial configuration is unperturbed, i.e. to choose the   
trivial solution\footnote{It is worth noting that this choice ensures that the Eulerian perturbation
of the induction equation (\ref{induction}) vanishes automatically, as required for 
an unconstrained variational formalism, see the discussion in \citet{CFS}.} $\Delta (B^a/\rho )=0 $ 
to (\ref{DB1}). This then leads to
\be
\Delta B^b = -B^b \nabla_c \xi^c
\label{DB2}
\ee
From this we readily obtain,
\be
\Delta B_a = \Delta g_{ab} B^b + g_{ab} \Delta B^b \quad \Rightarrow \quad 
\Delta B_a = B_b \nabla_a \xi^b -B_a\nabla_b \xi^b + B^b\nabla_b\xi_a
\label{DB3}
\ee 
The corresponding Eulerian perturbation of the magnetic field 
is given by 
\be
b^a \equiv \delta B^a = \Delta B^a - {\cal L}_{\xi} B^a \quad
\Rightarrow \quad b^a = -B^a (\nabla_b \xi^b) -(\xi^b \nabla_b) B^a
+ (B^b \nabla_b) \xi^a
\label{b_field}
\ee 
This is equivalent to \citep{roberts},
\be
b^a = \epsilon^{abc} \nabla_b \left ( \epsilon_{cde} \xi^{d} B^{e} \right )
\ee
where $\epsilon^{abc} $ is the familiar three-dimensional Levi-Civita symbol.
Note the useful identity,
\be
\epsilon^{abc} \epsilon_{ade} = \delta^{b}_{d}  \delta^{c}_{e}
-  \delta^{b}_{e}  \delta^{c}_{d}
\label{levi_id} 
\ee

%%%%%%%%%%%%%%%%%%%%%%%%%%%%%%%%%%%%%%%%%%%%%%%%%%%%%%%%%%%%%%%%%%%%%%%%%%

\subsection{Master equation}
\label{sec:master}

We now have all the tools required  to derive a `master' perturbation equation involving only the 
displacement vector $\xi^a$. This can be achieved by considering the Lagrangian perturbation of the Euler
equation~(\ref{euler}).
With the help of the background  equations  and after some rearrangement we  end up with 
\bear
&&\rho\,\partial_t^2 {\xi}_a + 2\rho u^b \nabla_b \partial_t {\xi}_a + \rho (u^b\nabla_b)^2 \xi_a
+ \delta\rho (\, \nabla_a \Phi + u^c\nabla_c u_a \,) + \rho \nabla_a \delta \Phi
+ \nabla_a \left ( \delta p + \frac{1}{4\pi} b^c B_c \right )
\nonumber \\
&& -\frac{1}{4\pi} (\, b^c \nabla_c B_a + B^c \nabla_c b_a \,) 
-\rho \xi^c \nabla_c ( u^b \nabla_b u_a ) =0
\label{master1}
\eear
This equation is of the form,
\be
A [\partial_t^2{\xi}_a] + B[\partial_t {\xi}_a] + C[\xi_a] =0
\label{master2} 
\ee
where $A,B,C$ are operators constructed from background variables. The first two
are identical to the operators appearing in the non-magnetic limit and, as shown by 
\citet{CFS}, they are respectively Hermitian and anti-Hermitian with respect 
to the inner product,
\be
\langle\, \eta^a, \xi_a\, \rangle \equiv \int (\eta^a)^\ast \xi_a dV
\label{inner}
\ee 
The integral is taken over the space occupied by the fluid and the asterisk
denotes complex a conjugate.    
That is, we have
\be
\langle\, \eta,A[\xi]\, \rangle = \langle\, \xi,A[\eta]\, \rangle^\ast ,\ \quad 
\langle\, \eta,B[\xi]\, \rangle = - \langle\, \xi,B[\eta]\, \rangle^\ast
\label{hermAB}
\ee
where both $\eta^a$ and $\xi^a$ are solutions of eqn.~(\ref{master2}).
Following \citet{CFS} we can construct the `symplectic' product,
\be
W(\eta,\xi) = \langle\, \eta,\Pi(\xi)\, \rangle - \langle\, \Pi(\eta),\xi\, \rangle
\ee
where we have introduced the conjugate canonical momentum,
\be
\Pi(\xi) = A [\partial_t{\xi}] + \frac{1}{2} B [\xi] = \rho \frac{D\xi}{Dt}
\ee
Then it follows (using eqns.~(\ref{master2}) and (\ref{hermAB})) that,
\be
\partial_t W(\eta,\xi) = \langle\, C[\eta],\xi\, \rangle -\langle\, \eta,C[\xi]\, \rangle
\label{Wdot}
\ee
This shows that, 
{\it provided} that the operator $C$ is Hermitian,  $W(\eta,\xi)$ is
a {\it conserved} quantity. This is  true for the non-magnetic case \citep{CFS}, and we can show that
it remains true for the magnetic case as well.

The proof of the Hermiticity of the $C$ operator is somewhat lengthy. In order to avoid 
overloading the reader
with algebra at this point, we relegate this  calculation to Appendix~\ref{app:hermitian}. 
The final result is, cf. Eq.~(\ref{Afinal}),
\bear
\eta^a C[\xi_a] &=& \nabla_a \left [ \eta^a ( \delta p + B^c b_c /4\pi )
-\frac{1}{4\pi} \eta^c ( B^a b_c + B^a \xi^b \nabla_b B_c )
+ \rho \eta^b u^a u^c \nabla_c \xi_b  
+ \delta_{\xi} \Phi \left (  \frac{1}{4\pi G} \nabla^a \delta_{\eta} \Phi
+ \rho \eta^a  \right )  \right ]  
\nonumber \\
&& -\frac{1}{4\pi G} \nabla^b \delta_{\eta} \Phi \nabla_b \delta_{\xi} \Phi
+ \frac{1}{4\pi} B^b B^c \nabla_c \eta^a \nabla_b \xi_a + \eta^a \xi^b 
\nabla_a \nabla_b ( p + B^2/8\pi ) + \nabla_a \eta^a \nabla_b \xi^b 
[ \gamma p + B^2/4\pi ]
\nonumber \\
&& + \frac{1}{4\pi} \left [ \nabla_a \eta^a B^c ( \xi^b \nabla_c B_b -\nabla_c \xi^b B_b )
+\nabla_b \xi^b B^c ( \eta^a \nabla_c B_a  -\nabla_c \eta^a B_a ) \right ]
\nonumber \\
&& -\rho ( u^c \nabla_c u_a + \nabla_a \Phi ) 
( \eta^a \nabla_b \xi^b + \xi^a \nabla_b \eta^b ) 
+ \rho \eta^a \xi^b \nabla_a \nabla_b \Phi -\rho u^c u^b \nabla_b \eta^a
\nabla_c \xi_a
\label{C1}
\eear
The fact that  $\eta^a$ and $\xi^a$ appear in a symmetric way proves that the operator $C$ is Hermitian,  
{\it modulo} the surface terms (resulting from the total divergence in the first line of (\ref{C1})). 
To discuss these remaining terms we write the result as,
\be
\eta^a C[\xi_a] = \eta^a C_\mathrm{H}[\xi_a] + \nabla_a S_\mathrm{M}^a + \nabla_a S^a_\mathrm{F} 
\label{C2}
\ee
where $C_\mathrm{H}$ is Hermitian and the surface terms are given by,
\bear
S_\mathrm{F}^a (\eta,\xi) &\equiv& \rho \eta^b u^a u^c\nabla_c \xi_b + \delta_{\xi} \Phi 
\left ( \frac{1}{4\pi G} g^{ab} \nabla_b \delta_{\eta}\Phi + \rho \eta^a \right )
\label{SF}
\\
\nonumber \\
S_\mathrm{M}^a (\eta,\xi) &\equiv& \eta^a \left (\, \delta p + \frac{1}{4\pi} B^c b_c\, \right )
-\frac{1}{4\pi} \eta^c \left ( \, B^a b_c + B^a \xi^b\nabla_b B_c \, \right )
\label{SM}
\eear
In the $B \to 0$ limit eqn.~(\ref{C2}) reduces to the result of \citet{CFS}. 
In this case the fluid surface terms vanish identically  since we have $\rho=p=\nabla p=0 $ and $\Delta p=0 $ 
at the  surface. The remaining gravitational term $ g^{ab} \delta_{\xi} \Phi \nabla_b \delta_{\eta}\Phi $
can be shown to be Hermitian, see \citet{kirsty}, rendering the full $C$ Hermitian. Using the same 
arguments it follows that the $S_\mathrm{F}$ term is Hermitian also when the magnetic field is present. 
The remaining surface term $S_\mathrm{M}^a$ is, however, generically 
non zero. This is easy to appreciate since the magnetic pressure does not have to vanish 
at the fluid surface. A simple way to avoid problems would be to assume that the surface terms
vanish anyway, e.g. by only considering magnetic fields that are confined to the fluid. 
This is, however, a severe and not very realistic restriction. Instead, we will consider the 
general problem and prove that the remaining surface terms are Hermitian.  
This requires further manipulation and a  detailed discussion of the appropriate boundary conditions 
that need to be satisfied at the fluid/vacuum interface.

%%%%%%%%%%%%%%%%%%%%%%%%%%%%%%%%%%%%%%%%%%%%%%%%%%%%%%%%%%%%%%%%%%%%%%%%%%%%%%%%%%%%%%%%%%%%%%%%%%%%%%%%%%%%%%%%%

\subsection{Surface boundary conditions}
\label{sec:surface}

Since we are interested in studying general systems which have a fluid-vacuum interface, 
we need to consider the relevant boundary conditions at such a surface. 
In order to work out these conditions we first consider the normal vector 
to the surface and how it is affected by a Lagrangian perturbation. 
The Lie derivative plays a key role in this analysis and simplifies the problem considerably,
(compared to, for example, the analysis in \citet{roberts}). 
Let us identify the surface as a level set of a scalar function $f$. In the fluid it would be natural
to identify this function with the fluid pressure, but we also assume that the function can be extended
to the exterior domain in a smooth fashion (this extension facilitates the definition of derivatives of
$f$). We then require,
\be 
[\partial_t + {\cal L}_u]\,f = 0
\ee  
from which it is easy to show that,
\be
[\partial_t + {\cal L}_u]\nabla_a f = 0
\label{Lie_norm}
\ee
In other words, the gradient $\nabla_a f $ is constant in the fluid frame. We also see that we must have,
\be
[\partial_t + {\cal L}_u]\Delta f = 0
\ee 
The trivial solution to this equation is $\Delta f =0 $, which essentially means that a fluid element at the
original surface remains at the perturbed surface. The normal to the surface can obviously be taken to
be $n_a = \nabla_a f $, from which a unit normal vector can be constructed,
\be
\hat{n}_a = \frac{\nabla_a f}{N}, \quad \mbox{where} \quad N = (g^{ab} \nabla_a f \nabla_b f)^{1/2}
\ee
Given the above results, it is straightforward to show that
\be
[\partial_t + {\cal L}_u]\hat{n}_a = -\frac{\hat{n}_a}{N}[\partial_t + {\cal L}_u] N 
\ee
This shows that any change to the unit normal is parallel to the normal itself. Hence the
orthogonal projection (into the surface) is conserved by the flow. This leads to,
\be
{ D\hn^a \over Dt} = ( \hn^a \hn^b - g^{ab} ) \hn^c \nabla_b u_c  
\ee
which is the equation for the normal vector given by, for example, \cite{kovetz}. Perturbing this equation
leads to,
\be
\Delta \hn_a = \hn_a \hn_c \hn^b \nabla_b \xi^c
\label{Dn_unit}
\ee
In deriving this we have assumed that the background configuration is such that $\hn^a u_a = 0 $, that is, 
there is no local expansion or contraction. Equation (\ref{Dn_unit}) highlights a key difference between 
the normal vectors $\hn_a$ and $n_a $, since for the latter we have from (\ref{Lie_norm}),
\be
[\partial_t + {\cal L}_u ] \Delta n_a = 0  
\ee
It is natural to use the trivial solution
\be
\Delta n_a = 0
\label{Dn}
\ee  
for fluid elements on the surface.     
Then it immediately follows that 
\be
g^{ab} \Delta n_b = \Delta n^a + n^c \nabla_c \xi ^a + g^{ab} n^c \nabla_b\, \xi_c = 0
\ee
That is, 
\be
\Delta n^a = - n^c \nabla_c \xi^a - g^{ab} n^c \nabla_b\, \xi_c 
\ee

Having a mathematical representation of the perturbed surface at hand,
we can move on to the surface condition for the magnetic and electric fields. 
The general conditions can be found in standard textbooks, see for example, \citet{roberts}.
In describing them we will
use angular brackets to represent the step in a quantity as we pass from the 
fluid to the exterior. That is, for any quantity $Q$ we define,
\be
\langle Q \rangle \equiv Q^{\rm x} - Q
\ee
where the index `$\rm x$' designates an exterior vacuum quantity. The brackets denoting a jump across the
boundary should not be confused with the brackets used in the definition of the inner product (\ref{inner}). 
One can easily avoid mixing things up by  keeping in mind that the symplectic product takes two arguments.

From the fact that $B^a $ is divergence-free we immediately obtain,
\be
\hn^a \langle\, B_a\, \rangle = 0
\label{Bbc}
\ee 
Meanwhile, the Maxwell equation $ \epsilon^{abc} \nabla_b E_c = -(1/c)\partial_t B^a $ leads to,
\be
\epsilon^{abc} \hn_b \langle\, E_c\, \rangle = { 1 \over c} ( \hn^b u_b ) \langle\, B^a\, \rangle
\label{Ebc1}
\ee

The surface conditions (\ref{Bbc}) and (\ref{Ebc1}) are originally formulated in terms of the
unit normal vector $\hn^a $. However in view of the previous results (eqns.~(\ref{Dn_unit}) and (\ref{Dn})) 
the forthcoming analysis is greatly simplified if we rewrite\footnote{Another
way of seeing that the perturbed norm of the normal vector does not affect the result is to note that, when we perturb
(\ref{Ebc1}) then $\Delta N$ will multiply a quantity that vanishes since the boundary conditions also 
hold for the background.} the surface conditions in terms of
$n^a = N \hn^a$. 
Subsequently taking the Lagrangian variation of (\ref{Ebc1}), and using the fact that $n_a u^a=0$ 
for the background, as well as eqn.~(\ref{Dn}), we find 
\be
c\, \hn_b \Delta ( \epsilon^{abc} \langle\, E_c\, \rangle ) = \hn_b \partial_t{\xi}^b \langle\, B^a\, \rangle 
\label{star}
\ee
Use of the background equation (\ref{ohm}) for the interior electric field 
together with  eqn.~(\ref{DB2}) allows us to show that
\be 
\Delta ( \epsilon^{abc} E_c ) = - { 1 \over c} \left [ ( \partial_t {\xi}^a B^b - \partial_t{\xi}^b B^a ) 
+ ( u^b B^a - u^a B^b ) \nabla_c \xi^c \right ]
\ee
For the exterior electric field we have
\be
\Delta ( \epsilon^{abc} E_c^{\rm x} ) = - \epsilon^{abc} E_c^{\rm x} \nabla_k \xi^k + \epsilon^{abc} \Delta E_c^{\rm x}
\ee 
where we have used \citep{CFS},
\be
 \Delta \epsilon^{abc} = -\epsilon^{abc} \,\nabla_k \xi^k
\label{levi}
\ee
Given these relations, eqn.~(\ref{star}) becomes
\be
c\,  \epsilon^{abc} \hn_b \Delta E_c^{\rm x} = \hn_k  ( \partial_t {\xi}^k B_{\rm x}^a - B^k \partial_t{\xi}^a )
+ u^a \hn_b B^b \nabla_k \xi^k   
\label{Ebc2}
\ee
where we have used (for the background)
\be
\epsilon^{abc} \hn_b E_c^{\rm x} = 0 
\ee
which follows from (\ref{Ebc1}).

Finally, it is useful to express the exterior electric field in terms of the vector potential
$ A^k$,
\be
E_k^{\rm x} = - { 1 \over c} \partial_t A_k 
\ee
Perturbing this we find
\be
\Delta E_k^{\rm x} = - { 1 \over c} \partial_t ( \Delta A_k - {\cal L}_\xi A_k ) =
- { 1 \over c} \partial_t \delta A_k \equiv - {1 \over c} \partial_t {\cal A}_k
\ee
where we have used the fact that the background is stationary.
This leads to the final form for the required boundary condition 
\be
\hn_b \partial_t{\xi}^b B_{\rm x}^a - \hn_b B^b \partial_t{\xi}^a + \epsilon^{abc} \hn_b \partial_t {\cal A}_c = 0
\ee
This relation is a total time derivative which upon integration gives
\be
\hn_b \xi^b B_{\rm x}^a - \hn_b B^b \xi^a + \epsilon^{abc} \hn_b {\cal A}_c = 0
\label{Ebc3}
\ee
This is identical to the condition derived by \citet{kovetz} in the case of a {\it static} background 
configuration.

An additional interface condition that needs to be satisfied is the continuity of the normal component of the fluid-magnetic 
field stress tensor,
\be
\langle\, T_{ab} \hn^b\, \rangle =0, \quad \mbox{where} \quad T_{ab} = -g_{ab}\left ( p + \frac{B^2}{8\pi} 
\right ) + \frac{1}{4\pi}\,B_a\,B_b
\label{stress}
\ee 
Explicitly,
\be
\hn_a \left \langle\,p + \frac{B^2}{8\pi}\, \right \rangle = \frac{1}{4\pi} (\hn_c B^c) \langle\, B_a \, \rangle
\ee
Projecting this along $\hn^a $ and using (\ref{Bbc}) we obtain,
\be
\left \langle\, p + \frac{B^2}{8\pi}\, \right \rangle = 0 
\label{pbc}
\ee
and
\be
( \hn^b B_b) \langle\, B_a\, \rangle = 0
\label{Bbc2}
\ee
Lagrangian variation of these conditions (replacing $\hn^a$ with $n^a $
in (\ref{Bbc2}), as before) results in,
\be
\Delta p = \frac{1}{4\pi} \left (\, B^{\rm x}_c b_{\rm x}^c + B^{\rm x}_c \xi^b \nabla_b B_{\rm x}^c
- B_c b^c + B_c \xi^b \nabla_b B^c  \, \right )
\label{Dp}
\ee
and
\be
\hn_c B^c \left (\, b^a -b_{\rm x}^a - \xi^b \nabla_b \langle\, B^a\, \rangle \,  \right ) = 0 
\label{DBbc2}
\ee

It is worth emphasising that we are assuming that ideal MHD can be used throughout our system. 
This assumption will not be valid for a realistic neutron star model since the Alfv\'en velocity will diverge 
near the fluid surface, as $\rho \to 0$. A detailed analysis of the surface region would require the introduction of
a transition region from ideal MHD conditions in the core of the star to the full Maxwell equations
in the exterior vacuum. Such an analysis is, however, much more intricate than our slightly inconsistent
approach. Once the stability problem has been completely understood within the present framework, it would 
be interesting to extend the analysis to a more detailed surface/vacuum model.

%%%%%%%%%%%%%%%%%%%%%%%%%%%%%%%%%%%%%%%%%%%%%%%%%%%%%%%%%%%%%%%%%%%%%%%%%%%%%%%%%

\subsection{Magnetic surface term}

\label{sec:surface_terms}

Returning to eqn.~(\ref{C2}), our  immediate objective is to eliminate, or symmetrise $S_\mathrm{M}$,  thus 
rendering the operator $C$ Hermitian. The following steps closely resemble the calculation by  \citet{kovetz}.

For convenience we split $S_\mathrm{M}$ into two pieces: $ \mathbf{\nabla}\cdot \mathbf{S}_\mathrm{M} = S_1 + S_2 $ with,
\bear
S_1 &=& \nabla_a \left [ \eta^a \left (  \delta p + \frac{1}{4\pi} B^c b_c \right )  \right ]
\label{S1}
\\
\nonumber \\
S_2 &=& -\frac{1}{4\pi}\, \nabla_a \left [ B^a \eta^c ( b_c + \xi^b \nabla_b B_c  ) \right ]
\label{S2}
\eear 

We write $\delta p = \Delta p -\xi^b \nabla_b p $, which with the help
of eqns.~(\ref{Dp}) and (\ref{S1}) leads to
\be
S_1 = \nabla_a \left [ \eta^a \left ( \frac{1}{4\pi} B_c^x b_x^c + \xi^b \nabla_b  
\left \langle\, p + \frac{B^2}{8\pi}\, \right  \rangle  \right  ) \right ]
\ee
Since eqn.~(\ref{pbc}) is uniformly valid everywhere at the fluid-vacuum interface we have
\be
\epsilon^{abc} \hn_b \nabla_c \left \langle\, p + \frac{B^2}{8\pi}\, \right \rangle = 0 
\ee
The cross product of this vector with $\xi^a $ is also zero, which leads to
%\be
%\epsilon^{abc} \xi_b \epsilon_{ckm} \hn^k g^{m\nu} \nabla_\nu \langle\, p + \frac{B^2}{8\pi}\, \rangle =0
%\ee
\be
\hn_a \xi^b \nabla_b \left \langle\, p + \frac{B^2}{8\pi}\, \right \rangle = 
\hn_b \xi^b \nabla_a \left \langle\, p + \frac{B^2}{8\pi}\, \right \rangle
\ee 

Hence,
\be
I_1(\eta,\xi) \equiv \int dV S_1 = \int dS \left ( \hn_a \hn_b \eta^a \xi^b \hn^c \nabla_c
\langle\, p + \frac{B^2}{8\pi}\, \rangle + \frac{1}{4\pi} \hn_a \eta^a B_c^x b^c_{\rm x} \right )
\ee
where the integral is taken over the space occupied by the fluid.
We can now utilise condition (\ref{Ebc3}) to obtain,
\be
I_1 = \int dS \left ( \hn_a \hn_b \eta^a \xi^b \hn^c \nabla_c
\left  \langle\, p + \frac{B^2}{8\pi}\, \right \rangle + \frac{1}{4\pi} \hn_b B^b \eta_c  b^c_{\rm x}
-\frac{1}{4\pi} \epsilon_{cmk} \hn^m {\cal A}^k b_{\rm x}^c \right )
\label{I1}
\ee 
In this expression the first term is symmetric with respect to $\eta$ and $\xi$ while the third
term involves only exterior fields. 

Turning to the second surface term $S_2 $ we have,
\be
I_2(\eta,\xi) \equiv \int dV S_2 = -\frac{1}{4\pi} \int dS \hn_a B^a \eta ^c \left ( b_c + \xi^b \nabla_b B_c \right )
\ee
Adding both integrals,
\bear
I_1 + I_2 &=& \int dS \left [ \hn_a \hn_b \eta^a \xi^b \hn^c \nabla_c
\left \langle\, p + \frac{B^2}{8\pi}\, \right  \rangle -\frac{1}{4\pi} \epsilon_{cmk} \hn^m {\cal A}^k b_x^c
+ \frac{1}{4\pi} \hn_a B^a \eta_c \left ( b^c_{\rm x} -b^c -\xi^b \nabla_b B^c  \right )  \right ]
\nonumber \\
&=& \int dS \left [ \hn_a \hn_b \eta^a \xi^b \hn^c \nabla_c
\left \langle\, p + \frac{B^2}{8\pi}\, \right \rangle -\frac{1}{4\pi} \epsilon_{cmk} \hn^m {\cal A}^k b_{\rm x}^c
- \frac{1}{4\pi} \hn_a B^a \eta^c \xi^b \nabla_b B^{\rm x}_c  \right ]
\eear
where we have used eqn.~(\ref{DBbc2}). 

Noting that $\epsilon^{abc} \nabla_b B^{\rm x}_c =0 \Rightarrow \nabla_b B_c^{\rm x} = \nabla_c B^{\rm x}_b $ we can symmetrise
the third term,
\be
\eta^c \xi^b \nabla_b B^{\rm x}_c = \frac{1}{2} \eta^c \xi^b ( \nabla_b B^{\rm x}_c + \nabla_c B^{\rm x}_b ) 
\ee

Finally, we can manipulate the `exterior' second term,
\be
\int dS\, \hn_m \epsilon^{cmk} {\cal A}_k b^{\rm x}_c = \int_{\rm x} dV\, \nabla_m 
\left ( \epsilon^{mkc} {\cal A}_k b^{\rm x}_c \right ) = 
\int_{\rm x} dV \left ( b^2_{\rm x} - {\cal A}_k \epsilon^{kmc} \nabla_m b_c^{\rm x}   \right )
\label{extint}
\ee
Note that the integration in the former two expressions is taken over the {\it exterior} vacuum. 
Since we generally assume $\epsilon^{abc} \nabla_b B_c = 0 $  in the exterior the second term in (\ref{extint}) 
vanishes. It it worth noting that \cite{kovetz} makes the same assumption. This is, of course, a
simplification since it means that we are not allowing electromagnetic waves. While this is consistent
with our approximate treatment of the surface, it would obviously be interesting to relax these 
constraints in a future analysis. The natural strategy for such a study would be to follow
the work of \citet{fried} who analyses the non-magnetic problem in full general relativity. 
The magnetic problem ought to be quite similar, with electromagnetic waves incorporated
essentially in the same way as gravitational waves.  
 
Collecting the above results, we arrive at
\be
I_\mathrm{M}(\eta,\xi) \equiv I_1 + I_2 = \int dS \left [ \hn_a \hn_b \eta^a \xi^b \hn^c \nabla_c
\left \langle\, p + \frac{B^2}{8\pi}\,\right \rangle -\frac{1}{8\pi} (\hn_a B^a) 
\eta^c \xi^b (\nabla_b B^{\rm x}_c + \nabla_c B^{\rm x}_b ) \right ] + 
\frac{1}{4\pi} \int_{\rm x} dV b_x^2
\ee
and clearly, $I_\mathrm{M}(\eta,\xi) = I_\mathrm{M}(\xi,\eta) $. 

Hence we have proven that the remaining magnetic surface term in (\ref{C2}) is Hermitian
\be
\int dV \nabla_a S_\mathrm{M}^a (\eta,\xi) = \int dV \nabla_a S_\mathrm{M}^a (\xi,\eta) 
\ee
This concludes the proof that the entire $C$ operator in (\ref{master2}) is Hermitian in the magnetic problem. 
It then follows from eqn.~(\ref{Wdot}) that the symplectic product $W(\eta,\xi)$ is a conserved quantity. 

%%%%%%%%%%%%%%%%%%%%%%%%%%%%%%%%%%%%%%%%%%%%%%%%%%%%%%%%%%%%%%%%%%%%%%%%%%%%%%%%%%%%%%%%%%%

\section{Canonical energy and angular momentum for magnetic stars}
\label{sec:energy}

As discussed by  \citet{CFS}, the symplectic product $W$ can be used to define the system's
canonical energy,
\be
E_c (\xi) \equiv \frac{1}{2} W(\partial_t{\xi},\xi) =
\frac{1}{2} \langle\, \partial_t{\xi},A[\xi]\, \rangle + \frac{1}{2} 
\langle\, \xi, C[\xi]\, \rangle
\label{Ec1}
\ee 
Written explicitly,
\be
E_c(\xi) = \frac{1}{2}\int dV \left ( \rho |\partial_t \xi |^2 + (\xi^a)^\ast C_\mathrm{H}[\xi_a] \right )
+ I_\mathrm{M}(\xi^\ast,\xi) 
\ee
or
\bear
E_c(\xi) &=& \frac{1}{2} \int dV \left \{ \,  \rho |\partial_t \xi |^2 -\rho | u^a \nabla_a \xi |^2
+ \left ( \gamma p + \frac{B^2}{4\pi} \right ) |\nabla_a \xi^a |^2 +
(\xi^a)^\ast \xi^b \left [ \rho \nabla_a \nabla_b \Phi + \nabla_a \nabla_b \left ( p + \frac{B^2}{8\pi} 
\right )  \right ] \right.
\nonumber \\
&-&  \left. 2\rho Re[(\xi^a)^\ast \nabla_b \xi^b ] \{ u^c \nabla_c u_a + \nabla_a \Phi  \} 
+ \frac{1}{4\pi} | B^a \nabla_a \xi |^2 + \frac{1}{2\pi} B^c 
Re \left \{ \nabla_a (\xi^a)^\ast ( \xi^b\nabla_c B_b -B_b \nabla_c \xi^b )  \right \}
\right.
\nonumber \\
&-& \left. \frac{1}{4\pi G}|\nabla \delta \Phi |^2 \right \}
+ \frac{1}{2} \int dS \left [ |\hn^a \xi_a |^2 \hn^c \nabla_c \left \langle\, p + \frac{B^2}{8\pi}\, \right \rangle  
-\frac{1}{8\pi} (\hn^a B_a) (\xi^c)^\ast \xi^b ( \nabla_b B^x_c + \nabla_c B^x_b ) \right ]
\nonumber \\
&+& \frac{1}{8\pi} \int_{\rm x} dV b_x^2
\label{Ec2}
\eear

A conserved canonical angular momentum $J_c$ can be defined in a similar fashion. 
We have
\be
J_c  \equiv -\frac{1}{2} W(\partial_\varphi {\xi},\xi) = 
- Re \langle \partial_\varphi, \Pi(\xi) \rangle
\ee
which is identical (as a functional) to the angular momentum expression of the non-magnetic problem \citep{CFS}. 
This is not surprising, since $J_c$ does not depend on the $C$ operator. 

It is useful to compare our result for $E_c$ to the corresponding expression for the non-magnetic
problem. The latter is simply obtained by setting $B^i=0$ in the above equations.  
We see that, in the magnetic case the canonical energy receives contributions from the fluid interior,
the vacuum exterior and the fluid-vacuum interface. This makes sense as this energy accounts for both 
fluid and magnetic-field degrees of freedom and the latter extend beyond the fluid body.

The canonical energy functional plays a central role in the formulation of a stability criterion for a
fluid configuration coupled to a radiative field. In its original form the criterion states that the configuration is stable
if $E_c(\xi) > 0  $ for {\it all} initial data \citep{bernstein,ostriker,CFS2}. 
In order for the system to be dynamically unstable we must have $E_c(\xi)= 0$, and
if $E_c(\xi) < 0 $ for {\it any} nonaxisymmetric data, then the configuration is secularly unstable. 
Intuitively, this criterion makes sense since the emitted radiation causes $\partial_t E_c < 0 $ invariantly. 
A perturbation which initially has negative canonical energy has the capability of growing in time while further 
decreasing its canonical energy. From our expression for the canonical energy (\ref{Ec2}) we learn that some of the terms 
associated with the magnetic degrees of freedom are positive definite, and thus act to stabilise the system.
At the same time, there are magnetic contributions to $E_c$ which are indefinite and which may (depending on the 
field configuration and the displacement vector) stimulate the onset of an instability.  
However, it would be premature to draw any conclusions about stability at this point. 
The above criterion is incomplete. Before applying it we must ensure that
the displacements we use correspond to a true physical perturbation.   

%%%%%%%%%%%%%%%%%%%%%%%%%%%%%%%%%%%%%%%%%%%%%%%%%%%%%%%%%%%%%%%%%%

\section{Trivial displacements}

\label{sec:trivials}

An important (and complicating) property of the Lagrangian perturbation theory for a fluid system is the existence
of a `gauge' freedom associated with the displacement vector $\xi^a$. As first discussed by \citet{CFS},
the theory admits a class of {\it trivial} displacements $\eta^a$ which correspond to vanishing Eulerian 
perturbations for all dynamical variables.  
Although the trivial displacements are unphysical they can have significant impact on the energy stability 
criterion \citep{CFS}. This is a consequence of the fact that trivial displacements\footnote{Note that for any trivial displacement 
$E_c (\eta) =0 $, provided the background configuration is {\it static} (see Appendix A of \citet{CFS}). 
In this case there is no issue of gauge dependence for the canonical energy.}   
are associated with a nonvanishing $E_c$.  
If we take any two displacements $\xi_1$ and $\xi_2 $ that differ by a 
trivial, i.e $\xi^a_2 = \xi^a_1 + \eta^a $, they obviously describe identical physical perturbations. 
Yet, $E_c (\xi_1 + \eta) \neq E_c(\xi_2)$ and it is easy to see how this could invalidate 
any stability conclusions drawn from the canonical energy. 
  
In order to make sense, the stability criterion must be reformulated in terms of `canonical'
displacements $\hxi$ which should have the property $E_c(\hxi + \eta) = E_c(\hxi) $. 
The family of canonicals is formally singled out by the requirement of orthogonality 
with respect to the product $W$,
\be
W(\hxi,\eta) = 0, \quad \mbox{for all trivials}~\eta
\label{canon}
\ee
It follows that if the canonicals  $\hxi_1$ and $\hxi_2 $ describe the same physical perturbation then
$\eta = \hxi_2 -\hxi_1 $ is a trivial and since both $\hxi_1$ and $\hxi_2 $ are orthogonal to all trivials, 
any trivial is itself orthogonal to all trivials, $W(\eta,\eta) = 0 $. Then clearly,
\be
E_c(\hxi_2) = \frac{1}{2}\, W(\partial_t \hxi_2,\hxi_2) = \frac{1}{2}\, W(\partial_t \hxi_1,\hxi_1)
+ \frac{1}{2}\, W(\partial_t \eta, \eta)  + W(\partial_t \hxi_1,\eta) = \frac{1}{2}\, W(\partial_t \hxi_1,\hxi_1)
= E_c(\hxi_1) 
\ee
Now that we understand the nature of the problem, it is clear that we need to investigate the class of trivial displacements
permitted in the magnetic problem. 

The general definition (\ref{canon}) leads to an explicit condition satisfied by every canonical displacement. 
In order to derive this condition we first need a solution for the trivials.     
In the fluid problem, the trivial diplacements $\eta^a$ are defined by the requirement
$\delta \rho = \delta u^a = 0$ which means,
\bear
&& \nabla_a (\rho \eta^a ) = 0
\label{triv1} \\
&& [ \partial_t + {\cal L}_u ] \eta^a = 0 
\label{triv2}
\eear
When a magnetic field is present we also require  $b^a = 0$ which provides
one additional equation
\be
\epsilon^{abc} \nabla_b \left ( \epsilon_{cde} \eta^{d} B^{e}  \right ) = 0
\label{triv3}
\ee
The remaining perturbations  $\delta p$ and $\delta E^a $ vanish as a consequence of the equation of state 
$p=p(\rho) $ and eqn.~(\ref{ohm}), respectively. Note that according to (\ref{triv2}) trivial displacements are 
always Lie-transported by the background flow.  In fact, the relation (\ref{triv2}) can be used to show that

\be
W(\hxi,\eta) = \int dV \rho \eta^a\, \Delta_{\hxi} u_a = 0 
\label{W1}
\ee

Before moving on to the magnetic problem, it is useful to recall what happens for a non-magnetic fluid.
Ignoring (\ref{triv3}), the solution of (\ref{triv1}) and (\ref{triv2}) is \citep{CFS}
\be
\eta^a = \frac{1}{\rho} \epsilon^{abc} \nabla_b \zeta_c
\label{triv_f}
\ee
where the vector field $\zeta_c$ is {\it arbitrary}, as long as it is Lie-transported
along the background flow,
\be
[ \partial_t + {\cal L}_u ] \zeta_a = 0
\label{zeta} 
\ee
This last result follows directly from $\delta u^a =0 $ and the identity 
\be
{\cal L}_u ( \nabla_a F_b ) = \nabla_a ( {\cal L}_u F_b ) 
- F_c \nabla_a \nabla_b u^c
\ee
Using (\ref{triv_f}) in (\ref{W1}) we obtain,
\be
\int dV\, 
\epsilon^{abc} \nabla_b (\Delta_{\hxi} u_a )\, \zeta_c = 0 
\ee
Since $\zeta_c$ is arbitrary, the above condition is satisfied if
\be
\epsilon^{abc} \nabla_b (\Delta_{\hxi} u_c) = 0 \quad \Rightarrow \quad 
\Delta_{\hxi} \left ( \frac{1}{\rho} \epsilon^{abc}\nabla_b u_c  \right ) = 0
\label{canon2}
\ee
The vector appearing in the last expression is the fluid's specific vorticity 
$\omega^a/\rho =  \epsilon^{abc}\nabla_b u_c/\rho$. Originally derived by \citet{CFS}, expression (\ref{canon2}) is the general condition which
should be satisfied by all canonical displacements. It states that a canonical displacement must preserve the 
vorticity of each fluid element. 

A different aspect of the same result can be revealed by taking the curl of the non-magnetic Euler equation (\ref{euler}).
This leads to the familiar vorticity equation,
\be
\frac{D\omega^a}{Dt} = \omega^b \nabla_b u^a -\omega^a \nabla_b u^b
\ee
This is easily written as,
\be
[\partial_t + {\cal L}_u ] \left (\frac{\omega^a}{\rho} \right) = 0 
\label{vort1}
\ee
which shows that the specific vorticity vector is Lie-transported. 
Perturbing (\ref{vort1}),
\be
[\partial_t + {\cal L}_u ] \Delta \left ( \frac{\omega^a}{\rho} \right ) = 0
\label{vort2}
\ee
Combining this general result with (\ref{canon2}) we see that the canonical displacements
single out the  the trivial solution $\Delta(\omega^a/\rho) = 0 $ of eqn.~(\ref{vort2}).

Finally, it is interesting to note that eqn.~(\ref{vort2}) can also be derived from 
$\partial_t W(\xi,\eta)=0 $ (where now $\xi $ is not necessarily canonical). 
Starting from (\ref{W1}) we can show that,
\be
W(\xi,\eta) = \int dV \zeta_c \rho \Delta \left (\frac{\omega^c}{\rho}  \right )
\ee
Taking the time derivative of this expression and making use of (\ref{zeta}) we find
\be
\partial_t W(\xi,\eta) = \int dV \zeta_c \rho\, [\partial_t + {\cal L}_u ]\Delta \left (\frac{\omega^c}{\rho} 
\right ) = 0
\label{Wdot1} 
\ee
Since $\zeta_c $ is arbitrary we end up with eqn.~(\ref{vort2}). 

Having arrived at the condition for a canonical displacement, we still face the issue of how 
to determine such displacements in practice. For later convenience, we note that
\be
\Delta_{\hxi} u_a = \nabla_a f
\label{canon3}
\ee 
where $f$ an arbitrary scalar function, is always canonical
\footnote{This can be immediately verified from (\ref{canon2}) for a non-magnetic fluid, and the same is true for
the case of a non-isentropic fluid \citep{CFS}.}. Indeed,
\be
W[\hat{\xi},\eta] = 0 \quad \Rightarrow \quad \int dV \nabla_a (\rho \eta^a ) f = 0
\ee
is identically satisfied due to (\ref{triv1}). 

Moving on to the magnetic problem, let us first provide an indication that it may be somewhat challenging. 
It is easy to see that the trivial form (\ref{triv_f}) with a general $\zeta_i$ cannot possibly be valid in the magnetic case. 
Indeed, if it were to remain true we could reproduce the derivation of (\ref{vort2}). 
However, in the MHD framework eqn.~(\ref{vort1})
is replaced by a more complicated relation,
\be
[\partial_t + {\cal L}_u ] \left (\frac{\omega^a}{\rho} \right) = \frac{1}{4\pi\rho} M^a
\ee 
with
\be
M^a = \epsilon^{abc}\, B^k \nabla_b B_k\, \nabla_c(1/\rho) +
 \epsilon^{abc} \nabla_b \left ( \frac{1}{\rho}\,B^k\,\nabla_k B_c \right )
\ee
Since the right-hand side of this equation is non-vanishing, we see that the
specific vorticity is not conserved in MHD. In fact, as far as we are aware, 
no corresponding conservation law is known. It is easy to see that this may 
complicate the analysis of the canonical displacements.

The equations that need to be solved in order to identify the class of trivial displacements
are (\ref{triv1}) and (\ref{triv3}). Since the general solution to the first of 
these equations is already known from the fluid problem we 'only' need to solve the magnetic 
equation and then link the two solutions in the appropriate way. 
First note that the general solution to (\ref{triv3}) can be written
\be
\epsilon_{cde} \eta^d B^e = \nabla_c \psi
\label{sol3}
\ee
where $\psi$ is an arbitrary scalar function\footnote{In terms of the magnetic vector potential $\mathbf{A}$ we have
\begin{displaymath}
\mathbf{b} = \mathbf{\nabla} \times \mathbf{\delta A} = \mathbf{\nabla} \times (\mathbf{\xi} \times \mathbf{B})
\end{displaymath}
which gives $\mathbf{\delta A} = \mathbf{\xi} \times \mathbf{B} + \mathbf{\nabla} f $ where $f$ represents the 
gauge degree of freedom. For a trivial displacement we have $\mathbf{\eta} \times \mathbf{B} = \mathbf{\nabla} \psi $
hence $\mathbf{\delta A} = \mathbf{\nabla} ( f + \psi ) $. Hence, the trivials in the magnetic case generate a 
pure gauge transformation for the magnetic vector potential. }.
This equation shows that
\be
\eta^a \nabla_a \psi = B^a \nabla_a \psi = 0
\ee
In other words, both the trivial displacement and the magnetic field lie in surfaces
of constant $\psi$.  
Taking the cross product of (\ref{sol3}) with $B^a$  we obtain
\be
B^2 \eta^a - (\eta^b B_b) B^a = \epsilon^{abc} B_b \nabla_c \psi 
\ee
Given this, it is easy to show that if we decompose the trivial into $\eta^a = \eta^a_\parallel + \eta^a_\perp$, where
$\eta^a_\parallel$ and $\eta^a_\perp$ are respectively parallel and orthogonal to $B^a$, then
\be
\eta^a_\perp = { 1 \over B^2} \epsilon^{abc} B_b \nabla_c \psi
\ee
We cannot, however, say anything about the parallel component at this point. 
Representing it by a scalar function $\alpha$, we have
\be
\eta^a = \alpha B^a + { 1 \over B^2} \epsilon^{abc} B_b \nabla_c \psi
\ee
This expression must also satisfy the fluid trivial condition (\ref{triv1}). This 
leads to a relation between $\alpha$ and $\psi$,  
\be
B^a \nabla_a (\rho \alpha) + \epsilon^{abc} \nabla_a \left( { \rho B_b \over B^2 }\right)(\nabla_c \psi)  = 0 
\label{constrain}\ee
Unfortunately, this relation shows that one should not expect to be able to completely determine $\alpha$ 
from a given $\psi$ (or vice versa). To see this note that the first term in (\ref{constrain}) is a directional derivative along $B^i$, 
while the second term is a scalar function (let us call it $g$). Suppose that we introduce a coordinate system
$[\hat{e}^i_1,\hat{e}^i_2, \hat{e}^i_3]$ such that $\hat{e}^i_1$ is aligned with $B^i$. Then (\ref{constrain})  
can be written
\be
B {\partial (\rho \alpha) \over \partial x_1} + g(x_1,x_2,x_3) = 0
\ee
which integrates to
\be
\alpha = - { 1 \over \rho} \int { g \over B} dx_1 + h(x_2,x_3)
\ee
where the function $h$ is freely specifiable. From this simple demonstration we learn that we need to treat the functions 
$\alpha$ and $\psi$ as (at least partially) independent.

A simple example of a case where $\alpha$ and $\psi$ are  independent is a
uniform density fluid and a uniform force-free magnetic field. Since $\epsilon^{abc} \nabla_b B_c = 0$ for
a force-free field and taking $\rho$ and $B^2$ as constants, the constraint (\ref{constrain}) collapses
to
\be
B^c \nabla_c (\rho \alpha) = 0
\ee 
from which it is clear that $\psi$ remains freely specifiable.

Even though we have failed to obtain an explicit expression for the trivial 
displacements in terms of a single arbitrary function,  we can proceed to write down the 
condition that determines a canonical displacement. Following the steps that led to 
(\ref{canon2}) we arrive at
\be
\left[ \rho \alpha B^a - \psi \epsilon^{abc} \nabla_c \left( {\rho B_b \over B^2} \right) \right]
\Delta_{\hxi} u_a - \psi \left( {\rho B_a \over B^2 }\right) \Delta_{\hxi} \left( { \omega^a \over \rho  } \right) = 0
\label{canons}\ee
Although this is not quite satisfactory, when combined with a solution to (\ref{constrain}) it does allow us to 
confirm whether a proposed perturbation is canonical or not. In fact, 
we already know that a class of canonical perturbations are given by $(\ref{canon3})$. It is easy to show that this 
class remains acceptable also in the magnetic case. 

As an example of applying (\ref{canons}) let us work out another 
set of canonical displacements. We see from (\ref{canons}) that the problem simplifies if we take
\be
B^a \Delta_{\hxi} u_a = 0
\ee
To satisfy (\ref{canons}) we additionally need (since $\psi$ is general)
\be
\nabla_a \left( { \rho \epsilon^{abc} B_c \Delta_{\hxi} u_b \over B^2} \right) = 0 
\ee 
One can show that this is satisfied by 
\be
\Delta_{\hxi} u_a = { 1 \over \rho} \left( B^c \nabla_a \chi_c - B^c \nabla_c \chi_a \right) 
\ee
for a general $\chi_a$.

%%%%%%%%%%%%%%%%%%%%%%%%%%%%%%%%%%%%%%%%%%%%%%%%%%%%%%%%%%%%%%%%%%%%%%%%%%%%%%%%%%%%%%%%%%%%%%%%%%%%%%%%%%%%%%%%%%

\section{Stability considerations}

\label{sec:stability}

Having at hand the canonical energy functional $E_c(\xi)$ and a properly formulated
criterion in terms of canonical diplacements, we are ready to investigate the 
stability of various model systems. As discussed by \citet{CFS} this analysis can be based on either normal modes
or canonical initial data. In the case of normal modes, i.e. solutions taking the form 
$\xi \sim e^{i\sigma t +im\varphi} $ (with $m$  the usual integer eigenvalue associated with the separation 
of the azimuthal $\varphi$ coordinate and $\sigma $ the frequency) one can show that they generally correspond to  
canonical displacements. Although there are exceptions to this rule\footnote{In the
case of a fluid configuration with  purely azimuthal flow $u^\varphi =\Omega(\varpi)$
a normal mode can belong to the family of trivials if it is corotating with the background flow, i.e
$\sigma = -m\Omega
$.} an instability proof based on a mode-solution is generally valid \citep{CFS}.  

Another important result that remains valid in the magnetic problem is the 
relation between the canonical energy and angular momentum 
\be
\frac{E_c}{J_c} = \omega_p \equiv -\frac{\sigma}{m}
\label{E_J}
\ee
This shows that the onset of instability can be deduced from a change in sign of 
the mode pattern speed $\omega_p$. 
If we imagine the star spinning up, an
initially counter-rotating mode with $J_c < 0$ becomes  unstable (in the sense of having 
negative $E_c$) beyond a critical rotation frequency.
The proof of the above relation proceeds exactly as in the non-magnetic problem 
since $E_c$ and $J_c $ have the same functional form  when expressed  in terms of the $A$ and $B$ operators of the master eqn.~(\ref{master2}). 
What {\it is} different between the magnetic and non-magnetic problems is the intrinsic character of the
modes themselves. A stability analysis for various classes of oscillation modes requires the 
determination of the associated displacement vectors. Solving for the oscillation modes of a rotating magnetised star is, however, 
a challenging problem. Once this problem is solved, the canonical energy formula 
can readily be used to test stability. 

In absence of actual mode-solutions for a rotating magnetic star, we 
can make progress by discussing various classes of initial data. This will provide some insight into
the role of the magnetic field. Let us start by extending the demonstration that a rotating star is generically
unstable due to the emission of radiation.  
In their discussion of the secular instability of (isentropic) fluids coupled to gravitational radiation, 
 \citet{CFS2} chose the following canonical data (in cylidrical coordinates $\varpi,z,\varphi$),
\bear
\hxi^a &=& (\mu \delta^a_\varphi + \nu^a) e^{im\varphi}
\label{dat1}
\\
\Delta_{\hxi} u_a &=& \nabla_a f \quad \Rightarrow \quad \partial_t \hxi_a = \nabla_a f -u^b (\nabla_b \hxi_a +
\nabla_a \hxi_b)
\label{dat2}
\eear  
with $f= \lambda e^{im\varphi} $. Given this choice we find,  
\bear
\nabla_b \xi^a &=& e^{im\varphi} \left [ im\delta_b^\varphi ( \mu \delta^a_\varphi + \nu^a ) + \delta^a_\varphi
\nabla_b \mu + \nabla_b \nu^a + \mu \left ( \frac{1}{\varpi} \delta^a_\varphi \delta^\varpi_b 
-\varpi \delta^a_\varpi \delta^\varphi_b \right )  \right ]
\label{var1}
\\
\nabla_a \xi^a &=& e^{im\varphi} ( im\mu + \nabla_k \nu^k)  
\label{var2}
\\
u^b \nabla_b \xi^a &=& e^{im\varphi} u^\varphi \left  [ im (\mu \delta^{a}_\varphi + \nu^a) -\mu \varpi \delta^a_\varpi + 
\frac{1}{\varpi} \nu^\varpi \delta^a_\varphi \right  ]
\label{var3}
\\
| u^b \nabla_b \xi^a |^2 &=&  (u^\varphi)^2 \left [ m^2 ( \varpi^2 \mu \mu^* + \nu^a \nu^*_a ) + 2im\varpi ( \mu \nu_\varpi^*
- \mu^* \nu_\varpi ) + \nu^\varpi \nu_\varpi^* + \varpi^2 \mu \mu^*  \right ]
\label{var4}
\\
|\partial_t \xi^a |^2 &=& m^2 \left [ \frac{1}{\varpi^2} \lambda\lambda^* + (u^\varphi)^2 ( 4\varpi^2 \mu \mu^*
+ X X^* + Z Z^*) -2 u^\varphi ( \mu \lambda^* + \mu^* \lambda )   \right ]
\label{var5}
\eear
where $Z = \nu^z $ and $ X = \nu^\varpi $. The resulting expression for the perfect fluid canonical energy, minimised with respect
to $\lambda $, is
\be
E_c \leq \, \frac{1}{2} m^2 \int dV \rho ( c_s^2 - \varpi^2 \Omega^2) \mu \mu^{*} + m k_1 + k_2
\ee
where $c_s^2 = \gamma p/\rho$ is the sound speed and the functions $k_1$ and $k_2 $ are independent of $m$. Since 
$c_s \to 0$ near the surface of the star, the energy functional becomes generically negative
in the 'multiarm' limit, $m \gg 1$.  Therefore, we can find physically acceptable perturbations that make 
{\it all} rotating fluid stars unstable to gravitational radiation.  
  
We can use the same initial data, eqns.~(\ref{dat1}) \& (\ref{dat2}), in the full MHD canonical energy (\ref{Ec2}). In addition
to (\ref{var1})-(\ref{var5}) we then need  the magnetic terms,
\bear
| B^c \nabla_c \xi^a |^2 &=& m^2 (B^\varphi)^2 ( \varpi^2 \mu \mu^* + X X^* + Z Z^* )
\\
B^c Re [ (\nabla_a \xi^a)^2 B_b \nabla_c \xi^b ] &=& m^2 B^\varphi \left \{ \varpi^2 \mu \mu^* B^\varphi
+ Re \left [ \mu ( B^\varpi X + B^z Z) \right ]  \right \}
\eear
This leads to
\bear
E_c &\leq & \frac{1}{2} m^2 \int dV \rho \left  \{ \mu \mu^{*} \left ( c^2_s -\varpi^2 \Omega^2
+ \frac{B_p^2}{4\pi\rho} \right ) +  \frac{(B^\varphi)^2}{4\pi\rho} ( X X^* + Z Z^* ) \right. 
\nonumber \\ 
&& \left. - \frac{B^\varphi}{4\pi\rho}\, [ \, B^\varpi ( \mu^{*} X + \mu X^{*}  )
+ B^z ( \mu^{*} Z + \mu Z^* ) \, ] \right \}  + m k_1 + k_2
\eear
where $B_p^2 = (B^\varpi)^2 + (B^z)^2 $. Since the first two magnetic terms are positive definite they
always play a stabilising role. Meanwhile, the third magnetic term can be either positive or negative.
Hence, it could potentially destabilise the system. It should be noted that this term vanishes when the 
magnetic field is either purely poloidal ($B^\varphi=0$) or purely toroidal ($B^\varpi=B^z=0$). 
It should, however, be present for all realistic neutron star field configurations.
In general the (negative) rotation term is the main destabilising factor. The main stabilising 
term is $\gamma p/\rho$
which is a surrogate for the star's gravitational potential energy. This energy is far greater than the energy
stored in the magnetic field and, consequently, provided we are not too close to the surface, the sound speed
term will dominate over any magnetic term (assuming components for the displacement vector $\mu,X,Z $ of roughly 
the same order of magnitude). Hence, one would expect the instability region to shift only slightly in the magnetic case. 
If a non-magnetic star is unstable at a given rotation rate for $m \gg 1$ then it is very likely that the
same is true for the corresponding magnetic star.   

To conclude the discussion, let us consider a  specific case  inspired by the instability of
the inertial r-modes in a rotating perfect fluid star \citep{na98,morsink}. Inertial modes are 
incompressible to leading order in the slow-rotation expansion. This means that it is interesting to 
consider displacements for which $\nabla_i \hxi^i=0$. In this case, the canonical energy simplifies considerably. 
Let us, for the sake of the argument, assume  an incompressible mode-solution of form (\ref{dat1}) 
with time-dependence represented by a frequency $\sigma$ (although it should be pointed out that it 
is not at all clear that such a solution actually exists). Then, further restricting the analysis to  a purely azimuthal 
magnetic field, we find in the $m \gg 1 $ limit,
\be
E_c \leq \frac{1}{2} \int dV \rho \left [ (g_{\varphi\varphi} \mu \mu^* + \nu^b \nu_b^*) \left ( \sigma^2 -m^2\Omega^2
+ m^2\frac{(B^\varphi)^2}{4\pi\rho} \right )  \right ] + m k_1 + k_2
\ee
This result is interesting because it suggests that in order for the mode to have a chance of becoming unstable 
the star's rotational frequency must first dominate over the Alfv\'en frequency of the background magnetic field.
This is a strong hint that the instability of (say) the r-modes may be significantly affected 
by the presence of a  magnetic field. This adds further complexity to the interesting issues discussed by 
\citet{rez1,rez2}. The problem of radiation driven instabilities in magnetic stars undoubtedly warrants further investigation.

%%%%%%%%%%%%%%%%%%%%%%%%%%%%%%%%%%%%%%%%%%%%%%%%%%%%%%%%%%%%%%%%%%%%%%%%%%%%%%%%%%%%%

\section{Concluding remarks}

In this paper we have presented a framework for analysing the stability properties of
a stationary and axisymmetric,
infinitely conducting perfect fluid configuration threaded by a magnetic field and surrounded by
vacuum. We have developed a Lagrangian perturbation approach which enables us to formulate a  rigorous stability criterion 
based on the canonical energy. 
For any  choice of initial data, represented by the displacement vector and its time derivative $\{\xi,\partial_t \xi  \} $, 
the sign of the canonical energy determines whether the 
configuration is stable or not at the linear perturbation level. 

Our analysis provides the first complete description of
the stability problem for a rotating magnetic star. We extended previous results for rotating stars by allowing for the presence
of an exterior vacuum field. We also discussed (for the first time in this context) the complications associated with the
so-called `trivial' fluid displacements, 
which do not represent true physical perturbations. In order for the stability criterion to make sense 
one has to isolate these trivials and consider only the physical `canonical' displacements. 
We discussed this problem and formulated a condition which must be satisfied by all canonical displacements.    
Having obtained a well-defined stability criterion we provided two examples that 
indicate that the magnetic field tends to stabilise radiation driven instabilities. 
 
To make further progress we need to calculate explicit mode-solutions for a rotating magnetic system.
Once such modes have been obtained the canonical energy formulated in this paper can be used to 
assess their stability. Of course, the calculation of oscillation modes of a magnetic star is a 
far from trivial problem that sets a severe challenge for future efforts in this area. On the other
hand, strong motivation for such work is provided by increasingly detailed observations of for example magnetar
flares and associated quasiperiodic oscillations.

%%%%%%%%%%% -- APPENDICES --  %%%%%%%%%%%%%%%%%%%%%%%%%%%%%%%%%%%%%%%%%%%%%%%%%%%

\appendix

\section{Proof of Hermiticity of the $C$ operator}
\label{app:hermitian}

In this Appendix we describe the various steps involved in proving the
Hermiticity of the $C $ operator.  
We start from (see the main text)
\bear
\eta^a C[\xi_a] &=& \rho \eta^a (u^b\nabla_b)^2 \xi_a -\eta^a \nabla_b(\rho\xi^b)
( \nabla_a \Phi + u^c \nabla_c u_a ) + \rho \eta^a \nabla_a \delta \Phi 
+ \eta^a \nabla_a \left ( \delta p + \frac{1}{4\pi} b^c B_c \right ) 
\nonumber \\
&& -\frac{1}{4\pi} \eta^a ( b^c \nabla_c B_a + B^c \nabla_c b_a ) -\rho \eta^a \xi^c
\nabla_c ( u^b \nabla_b u_a )
\eear 
or,
\bear
\eta^a C[\xi_a] &=& \nabla_a \left [\eta^a \left ( \delta p + \frac{1}{4\pi} b^c B_c 
\right ) \right ] + \rho \eta^a (u^b\nabla_b)^2 \xi_a -\eta^a \nabla_b(\rho\xi^b)
( \nabla_a \Phi + u^c \nabla_c u_a ) + \rho \eta^a \nabla_a \delta \Phi 
\nonumber \\
&&- \nabla_a \eta^a \left ( \delta p + \frac{1}{4\pi} b^c B_c \right ) 
-\frac{1}{4\pi} \eta^a ( b^c \nabla_c B_a + B^c \nabla_c b_a ) -\rho \eta^a \xi^c
\nabla_c ( u^b \nabla_b u_a )
\eear 
Our task is to manipulate $\eta^a C[\xi_a] $  to make it symmetric
with respect to $\eta^a$ and $\xi^a$. 

Let us first study separately the magnetic `tension' term,
\bear 
-\frac{1}{4\pi} \eta^a ( b^c \nabla_c B_a + B^c \nabla_c b_a ) &=& \frac{1}{4\pi}
\eta^a ( \xi^b \nabla_b B^c \nabla_c B_a - B^b \nabla_b \xi^c \nabla_c B_a
+ \nabla_b \xi^b B^c \nabla_c B_a ) -\frac{1}{4\pi} \nabla_c (\eta^a B^c b_a )
\nonumber \\
&& + \frac{1}{4\pi} B^c \nabla_c \eta^a ( -\nabla_b \xi^b B_a - \xi^b\nabla_b B_a
+ B^b \nabla_b \xi_a  )
\nonumber \\
\nonumber \\
&=&  -\frac{1}{4\pi} \nabla_c ( \eta^a B^c b_a ) + \frac{1}{4\pi} B^c B^b \nabla_c 
\eta^a \nabla_b \xi_a -\frac{1}{4\pi} \nabla_b \xi^b B^c ( B_a \nabla_c \eta^a
- \eta^a \nabla_c B_a  ) 
\nonumber \\
&& - \frac{1}{4\pi} ( -\eta^a \xi^b \nabla_b B^c \nabla_c B_a
+ \eta^a B^b \nabla_b \xi^c \nabla_c B_a + B^c \nabla_c \eta^a \xi^b \nabla_b B_a )
\eear
And since,
\be
\eta^a B^b \nabla_b \xi^c \nabla_c B_a = \nabla_b ( \eta^a B^b \xi^c \nabla_c B_a )
- \nabla_b \eta^a B^b \xi^c \nabla_c B_a -\eta^a \xi^c B^b \nabla_b \nabla_c B_a
\ee
we have,
\bear
-\frac{1}{4\pi} \eta^a ( b^c \nabla_c B_a + B^c \nabla_c b_a ) &=& -\frac{1}{4\pi}
\nabla_c ( \eta^a B^c b_a + \eta^a B^c \xi^b \nabla_b B_a ) +\frac{1}{4\pi}
B^c B^b \nabla_c \eta^a \nabla_b \xi_a 
\nonumber \\
&& - \frac{1}{4\pi} \nabla_b \xi^b B^c 
( B_a \nabla_c \eta^a -\eta^a \nabla_c B_a ) + \frac{1}{4\pi} \eta^a \xi^b
\nabla_b ( B^c \nabla_c B_a )
\eear
Using once the background Euler equation in the last term,
we arrive at,
\bear
-\frac{1}{4\pi} \eta^a ( b^c \nabla_c B_a + B^c \nabla_c b_a ) &=& -\frac{1}{4\pi}
\nabla_c ( \eta^a B^c b_a + \eta^a B^c \xi^b \nabla_b B_a ) + \frac{1}{4\pi}
B^b B^c \nabla_c \eta^a \nabla_b \xi_a 
\nonumber \\
&&- \frac{1}{4\pi} \nabla_b \xi^b B^c 
( B_a \nabla_c \eta^a -\eta^a \nabla_c B_a ) 
+ \eta^a \xi^b \nabla_a \nabla_b \left ( p + \frac{B^2}{8\pi} \right )
\nonumber \\
&&+ \eta^a \xi^b \nabla_b [ \rho ( u^c \nabla_c u_a + \nabla_a \Phi) ]
\eear
Moreover, we have for the total pressure term,
\bear
-\nabla_a \eta^a \left ( \delta p + \frac{1}{4\pi} b^c B_c \right ) &=&
\nabla_a \eta^a ( \gamma p \nabla_b \xi^b + \xi^b \nabla_b p )
-\frac{1}{4\pi} \nabla_a \eta^a B_c ( -\nabla_b \xi^b B^c -\xi^b \nabla_b B^c
+ B^b \nabla_b \xi^c )
\nonumber \\ 
&=& \nabla_a \eta^a \nabla_b \xi^b ( \gamma p + B^2/4\pi ) 
+ \nabla_a \eta^a \xi^b \nabla_b ( p + B^2/8\pi )
-\frac{1}{4\pi} \nabla_a \eta^a B^b \nabla_b \xi^c B_c
\eear
Using once more the background Euler equation in the second term we obtain,
\bear
-\nabla_a \eta^a \left ( \delta p + \frac{1}{4\pi} b^c B_c \right ) &=&
\nabla_a \eta^a \nabla_b \xi^b ( \gamma p + B^2/4\pi )
-\rho \nabla_a \eta^a \xi^b ( \nabla_b \Phi + u^c \nabla_c u_b  )
\nonumber \\
&& + \frac{1}{4\pi} \nabla_a \eta^a B^c ( \xi^b \nabla_c B_b
- \nabla_c \xi^b B_b )
\eear
Collecting the above results and returning to the full expression,
\bear
\eta^a C[\xi_a] &=& \nabla_a \left [ \eta^a \left ( \delta p + \frac{1}{4\pi}B^c b_c 
\right ) -\frac{1}{4\pi} ( \eta^c B^a b_c + \eta^c B^a \xi^b \nabla_b B_c  ) \right ]
+ \frac{1}{4\pi} B^b B^c \nabla_c \eta^a \nabla_b \xi_a
\nonumber \\
&& + \eta^a \xi^b \nabla_a \nabla_b ( p + B^2/8\pi ) 
+ \nabla_a \eta^a \nabla_b \xi^b ( \gamma p + B^2/4\pi ) 
+ \rho \eta^a \nabla_a \delta \Phi
\nonumber \\
&& + \frac{1}{4\pi} \left [ \nabla_a \eta^a B^c ( \xi^b \nabla_c B_b 
-\nabla_c \xi^b B_b )
-\nabla_b \xi^b B^c ( B_a \nabla_c \eta^a -\eta^a \nabla_c B_a 
) \right ]  -\eta^a \nabla_b (\rho \xi^b) ( \nabla_a \Phi + u^c \nabla_c u_a )
\nonumber \\
&& -\rho \eta^a \xi^c \nabla_c ( u^b \nabla_b u_a ) + \rho \eta^a (u^b \nabla_b)^2 \xi_a
+ \eta^a \xi^b \nabla_b [\rho ( u^c \nabla_c u_a + \nabla_a \Phi ) ]
-\rho \nabla_a \eta^a \xi^b ( \nabla_b \Phi + u^c \nabla_c u_b )
\label{nCxi}
\eear
The remaining non-symmetric terms (apart from the first surface term) 
are the last five terms. Denoting them collectively as $R$ we have,
\bear
R &=& -\eta^a \nabla_b (\rho \xi^b) ( \nabla_a \Phi + u^c \nabla_c u_a )
-\rho \eta^a \xi^c \nabla_c ( u^b \nabla_b u_a ) + \rho \eta^a (u^b \nabla_b)^2 \xi_a
\nonumber \\
&& + \eta^a \xi^b \nabla_b [\rho ( u^c \nabla_c u_a + \nabla_a \Phi ) ]
-\rho \nabla_a \eta^a \xi^b ( \nabla_b \Phi + u^c \nabla_c u_b )
\eear 
Differentiating by parts the fourth term we find,
\be
R = -\rho ( u^c \nabla_c u_a + \nabla_a \Phi ) ( \eta^a \nabla_b \xi^b
+ \xi^a \nabla_b \eta^b ) + \rho \eta^a \xi^b \nabla_a \nabla_b \Phi
+ \rho \eta^a u^b \nabla_b ( u^c \nabla_c \xi^a )
\ee 
Furthermore,
\bear
\rho \eta^a u^b \nabla_b ( u^c \nabla_c \xi^a ) &=& \rho \eta^a u^b \nabla_b u^c 
\nabla_c \xi_a + \rho \eta^a u^b u^c \nabla_b \nabla_c \xi_a
\nonumber \\
&=& \nabla_b ( \rho \eta^a u^b u^c \nabla_c \xi_a ) -\rho \nabla_b\eta^a
u^b u^c \nabla_c \xi_a -\eta^a u^c \nabla_c\xi_a ( u^b \nabla_b \rho 
+ \rho \nabla_b u^b ) 
\eear
The last term vanishes due to the background continuity equation so eventually we are left with,
\bear
R &=& \nabla_b ( \rho \eta^a u^b u^c \nabla_c \xi_a ) 
-\rho ( u^c \nabla_c u_a + \nabla_a \Phi ) 
( \eta^a \nabla_b \xi^b + \xi^a \nabla_b \eta^b )
\nonumber \\
&& + \rho \eta^a \xi^b \nabla_a \nabla_b \Phi 
-\rho u^c u^b \nabla_b \eta^a
\nabla_c \xi_a
\eear
Written in this form, $R$ contains only surface and symmetric terms.
The final step is to symmetrise the gravitational potential term 
$\rho\, \eta^a \nabla_a \delta \Phi$ in eqn.~(\ref{nCxi}). 
From Ref.~\cite{CFS} we have,
\bear
\rho\, \eta^a \nabla_a \delta \Phi &=& \nabla_a ( \rho \eta^a \delta_{\xi} \Phi )
+ \delta_{\eta} \rho \delta_{\xi} \Phi
\nonumber \\
&=& \nabla_a (\rho \eta^a \delta_{\xi} \Phi ) + \frac{1}{4\pi G} \nabla^2 \delta_{\eta} \Phi
\delta_{\xi} \Phi
\nonumber \\
&=& \nabla_a \left [ \delta_{\xi} \Phi \left (  \frac{1}{4\pi G} \nabla^a \delta_{\eta} \Phi
+ \rho \eta^a  \right ) \right ] -\frac{1}{4\pi G} \nabla^b \delta_{\eta} \Phi
\nabla_b \delta_{\xi} \Phi
\eear 
Collecting all the  results we finally obtain,
\bear
\eta^a C[\xi_a] &=& \nabla_a \left [ \eta^a ( \delta p + B^c b_c /4\pi )
-\frac{1}{4\pi} \eta^c ( B^a b_c + B^a \xi^b \nabla_b B_c )
+ \rho \eta^b u^a u^c \nabla_c \xi_b  
+ \delta_{\xi} \Phi \left (  \frac{1}{4\pi G} \nabla^a \delta_{\eta} \Phi
+ \rho \eta^a  \right )  \right ]  
\nonumber \\
&& -\frac{1}{4\pi G} \nabla^b \delta_{\eta} \Phi \nabla_b \delta_{\xi} \Phi
+ \frac{1}{4\pi} B^b B^c \nabla_c \eta^a \nabla_b \xi_a + \eta^a \xi^b 
\nabla_a \nabla_b ( p + B^2/8\pi ) + \nabla_a \eta^a \nabla_b \xi^b 
( \gamma p + B^2/4\pi )
\nonumber \\
&& + \frac{1}{4\pi} [ \nabla_a \eta^a B^c ( \xi^b \nabla_c B_b -\nabla_c \xi^b B_b )
+\nabla_b \xi^b B^c ( \eta^a \nabla_c B_a  -\nabla_c \eta^a B_a ) ]
\nonumber \\
&& -\rho ( u^c \nabla_c u_a + \nabla_a \Phi )
( \eta^a \nabla_b \xi^b + \xi^a \nabla_b \eta^b ) 
+ \rho \eta^a \xi^b \nabla_a \nabla_b \Phi -\rho u^c u^b \nabla_b \eta^a
\nabla_c \xi_a
\label{Afinal}
\eear
This is eqn. (\ref{C1}) of the main text.

%%%%%%%%%%%%%%%%%%%%%%%%%%%%%%%%%%%%%%%%%%%%%%%%%%%%%%%%%%%%%%%%%%%%%%%%%%%%%%%%%%%%%%%%%

\section*{Acknowledgments}

This work was supported by PPARC through grant number PPA/G/S/2002/00038.
NA  acknowledges support from PPARC via Senior Research Fellowship no
PP/C505791/1. We would also like to thank the anonymous referee for his/her 
constructive criticism that led to significant improvements of the paper.

%%%%%%%%%%%%%%%%%%%%%%%%%%%%%%%%%%%%%%%%%%%%%%%%%%%%%%%%%%%%%%%%%%%%%%%%%%%%%%%%%%%

%%%%%%%%%%%%%%%%

\end{document}